\documentclass[sigconf]{acmart}

\AtBeginDocument{%
  }

\setcopyright{acmlicensed}
\copyrightyear{2018}
\acmYear{2018}
\acmDOI{XXXXXXX.XXXXXXX}

\acmConference[Conference acronym 'XX]{Make sure to enter the correct
  conference title from your rights confirmation emai}{June 03--05,
  2018}{Woodstock, NY}
\acmISBN{978-1-4503-XXXX-X/18/06}

\usepackage{tikz}
\usepackage{amsmath}

\usepackage{xcolor,color,xspace,enumerate,centernot,multirow,float,graphicx,
xcolor,caption,subcaption,textcomp,pgfplots,pgf-pie,tikz,listings,enumitem,
comment,adjustbox,mdframed,changepage,algorithm,algorithmic,soul,msc,tablefootnote}
\usepackage{hyperref}
\usepackage{filecontents}
\usepackage{xurl} % for usenix
\usepackage{makecell}
\usepackage{comment}

\usepackage[utf8]{inputenc}
\usetikzlibrary{shapes}
\usetikzlibrary{arrows.meta}
\usetikzlibrary{arrows}
\usetikzlibrary{positioning}
\usetikzlibrary{shadows}
\usetikzlibrary{backgrounds}
\usepackage[most]{tcolorbox}
\definecolor{lightblue}{rgb}{0.68, 0.85, 0.9}
\definecolor{lightpink}{rgb}{1.0, 0.71, 0.76}
  {\end{adjustwidth}}

\definecolor{darkgray}{gray}{0.6}

\newlist{Qenumerate}{enumerate}{1}
\setlist[Qenumerate]{label={Q}\arabic*}
\newlist{Cenumerate}{enumerate}{1}
\setlist[Cenumerate]{label={C}\arabic*}

\newcommand{\liboqs}{\textbf{\texttt{liboqs}}}
\newcommand{\circl}{\textbf{\texttt{CIRCL}}}
\newcommand{\awslc}{\textbf{\texttt{AWS-LC}}}
\newcommand{\pqm}{\textbf{\texttt{pqm4}}}

\newcommand{\pqclean}{\textbf{\texttt{PQClean}}}
\newcommand{\libpq}{\textbf{\texttt{libpqcrypto}}}

\AtBeginDocument{%
  }

\date{}

\title{Study of Post Quantum status of Widely Used Protocols}

\author{Tushin Mallick}
\authornote{This work was carried out when the author was at Cisco Research, San Jose, CA, during Summer Internship 2025. He is currently at Northeastern University, Boston, MA.}
\affiliation{
  \institution{Cisco Research}
    \city{}
  \country{}
}

\author{Ashish Kundu}
\affiliation{
  \institution{Cisco Research}
  \city{}
  \country{}
}
\author{Ramana Kompella}
\affiliation{
  \institution{Cisco Research}
  \city{}
  \country{}
}
\begin{document}

\begin{abstract}
The advent of quantum computing poses significant threats to classical public-key cryptographic primitives such as RSA and elliptic-curve cryptography. As many critical network and security protocols depend on these primitives for key exchange and authentication, there is an urgent need to understand their quantum vulnerability and assess the progress made towards integrating post-quantum cryptography (PQC). This survey provides a detailed examination of nine widely deployed protocols ---TLS, IPsec, BGP, DNSSEC, SSH, QUIC, OpenID Connect, OpenVPN, and Signal Protocol ---analysing their cryptographic foundations, quantum risks, and the current state of PQC migration. We find that TLS and Signal lead the transition with hybrid post-quantum key exchange already deployed at scale, while IPsec and SSH have standardised mechanisms but lack widespread production adoption. DNSSEC and BGP face the most significant structural barriers, as post-quantum signature sizes conflict with fundamental protocol constraints. Across all protocols, key exchange proves consistently easier to migrate than authentication, and protocol-level limitations such as message size and fragmentation often dominate over raw algorithm performance. We also discuss experimental deployments and emerging standards that are shaping the path towards a quantum-resistant communication infrastructure.
\end{abstract}

\maketitle

\section{Introduction}
Quantum computing uses quantum mechanics to solve certain problems exponentially faster than classical computers. In particular, Shor's algorithm breaks the integer factorisation and discrete logarithm problems that underpin RSA, Diffie–Hellman, and elliptic-curve cryptography. Although large-scale quantum computers do not yet exist, adversaries can already record encrypted traffic today and decrypt it once quantum capabilities mature ---the so-called ``harvest-now, decrypt-later'' threat. For data requiring long-term confidentiality, this renders current protections inadequate well before a quantum computer is physically built.

NIST conducted a multi-year effort to standardise quantum-resistant algorithms, releasing public drafts in 2023~\cite{nist-pqc-details,nist-fips203_encryption,nist-fips204_ds,nist-fips205_hds} and publishing the first PQC standards ---FIPS~203 (ML-KEM), FIPS~204 (ML-DSA), and FIPS~205 (SLH-DSA) ---in August 2024~\cite{NIST-pqc-standard-Aug132024,NIST-pqc-standard-fips203,NIST-pqc-standard-fips204,NIST-pqc-standard-fips205}. In March 2025, NIST selected HQC as a fifth key-encapsulation mechanism, with a draft standard expected in 2026 and finalisation by 2027. Draft NIST IR~8547 recommends deprecating quantum-vulnerable algorithms (RSA, ECDSA, EdDSA, DH, ECDH) by 2030 and fully retiring them by 2035.

PQC primitives are now available in production-grade libraries. \liboqs~\cite{liboqs}, from the Open Quantum Safe project, supports integration and evaluation of PQC algorithms across major operating systems. \circl~\cite{circl} provides PQC primitives for the Go language. Amazon's \awslc~\cite{aws-lc} incorporates PQC into AWS security services. For constrained environments, \pqm~\cite{pqm4} targets ARM Cortex-M4 platforms. \pqclean~\cite{pqclean} offers portable C reference implementations of NIST candidates, and \libpq~\cite{libpqcrypto} collects diverse schemes for research and benchmarking.

However, algorithm availability alone does not imply protocol readiness. Integrating PQC into deployed protocols involves increased key and signature sizes, fragmentation and round-trip overhead, backward compatibility constraints, and a lack of operational experience at scale. Modern systems rarely depend on a single protocol: a typical enterprise deployment relies on TLS for web traffic, IPsec for site-to-site VPNs, SSH for remote administration, DNSSEC for name resolution integrity, BGP for routing security, QUIC for low-latency transport, and OIDC for federated authentication. Each of these protocols embeds classical cryptography in protocol-specific ways, and each faces distinct migration challenges. Yet existing studies tend to examine post-quantum migration for individual protocols in isolation, leaving practitioners and policymakers without a unified view of where the transition stands, which protocols are furthest along, and where the critical bottlenecks lie.

This paper addresses that gap by surveying nine widely deployed protocols --- TLS, IPsec, BGP, DNSSEC, SSH, QUIC, OpenID Connect, OpenVPN, and the Signal Protocol. For each, we analyse its current cryptographic reliance, quantum vulnerabilities, post-quantum migration efforts to date, and the challenges that remain. Our goal is to provide a single, cross-protocol reference that captures the state of the post-quantum transition as of 2025, enabling informed migration planning across the protocol stack.

\section{Background}
\label{sec:pq}

PQ cryptography is crucial to ensuring security in the face of emerging quantum computing threats. As quantum computers advance, they pose a risk to classical cryptographic algorithms, making the adoption of PQ key exchange mechanisms and digital signature schemes essential. Gradually, these PQ algorithms are being integrated into existing protocols to enhance security and future-proof systems against quantum attacks. Below
we overview current PQ algorithms and some protocols that implement them.

\subsection{KEM and Digital Signature Algorithms}

The algorithms are categorized based on the mathematical problems they rely on, each offering different strengths, weaknesses, and applications. A summary of the algorithms is provided in Table \ref{tab:postquantum} and we expound on the categories below. 

\textbf{Lattice-based} cryptography is a leading class of PQC, grounded in hard lattice problems such as the Shortest Vector Problem and Learning With Errors, which are believed to resist both classical and quantum attacks. These schemes offer favorable efficiency and compactness, making them well suited for key exchange, encryption, and digital signatures. Notable examples include the key-exchange mechanisms CRYSTALS-Kyber \cite{kyber2018}, NTRU \cite{ntru1998}, and SABER \cite{saber2018}, as well as the digital signature scheme CRYSTALS-Dilithium, all of which are finalists in the NIST PQC standardization process.

\textbf{Code-based} cryptography is rooted in the hardness of decoding random linear codes—a problem considered difficult even for quantum computers.  While this category of algorithms offer small ciphertexts and fast encryption/decryption, their main drawback is the large size of public keys, which can be challenging to manage. Key exchange algorithm Classic McEliece\cite{mceliece1978} is a finalist in the NIST PQC standardization process for this category.

\textbf{Hash-based} cryptography derives its security solely from cryptographic hash functions, making it especially well suited for digital signatures. Unlike lattice-based schemes, it does not depend on complex mathematical structures but on the hardness of finding hash collisions, a problem for which quantum computers offer only limited advantage. While providing strong security guarantees, hash-based schemes generally incur larger signature sizes.

\textbf{Multivariate polynomial} cryptography is based on the hardness of solving systems of multivariate quadratic equations over finite fields, an NP-hard problem. Schemes such as Rainbow \cite{rainbow2005} and GeMSS \cite{gemss2018} are mainly designed for digital signatures and offer PQ security, albeit with large key and signature sizes. Rainbow is a finalist in the NIST PQC standardization process for this category.

\textbf{Isogeny-based} cryptography exploits the difficulty of computing isogenies between elliptic curves and is notable for compact key sizes, making it attractive for constrained environments. However, in July 2022, KU Leuven researchers broke the SIKE \cite{sike} algorithm, a NIST fourth-round candidate, using a classical computer in 62 minutes, highlighting vulnerabilities in its underlying supersingular isogeny problem.

\textbf{Zero-knowledge(ZK)} proof-based PQ algorithms are designed to enable secure verification processes without revealing the underlying data, even in the face of quantum computing threats. 

\textbf{Hybrid Algorithms.} PQ hybrid algorithms combine conventional cryptography with PQC to ensure security during the transition to a quantum-resistant era. They generate parallel key pairs---one from a classical scheme (e.g., RSA or ECC) and one from a PQ scheme---so that security is preserved as long as at least one component remains unbroken. Consequently, hybrid approaches provide robust, forward-looking protection and are expected to play a central role in securing sensitive data against both classical and quantum adversaries.

\textbf{Standardization.} 
In 2022, NIST selected CRYSTALS-Kyber, FALCON, SPHINCS+ and CRYSTALS-Dilithium for standardization, releasing draft standards for three in 2023. In August 2024, NIST published the first PQC standards~\cite{NIST-pqc-standard-Aug132024}: FIPS~203~\cite{NIST-pqc-standard-fips203}, FIPS~204~\cite{NIST-pqc-standard-fips204}, and FIPS~205~\cite{NIST-pqc-standard-fips205}. These finalize Kyber as ML-KEM (key encapsulation), Dilithium as ML-DSA (digital signatures), and SPHINCS+ as SLH-DSA (stateless hash-based signatures). NIST is developing a second set of standards, including FIPS~206 based on FALCON (FN-DSA); however, as of 2025, no final release has been announced.

\begin{table*}[h!]
\centering
\scriptsize
\setlength{\tabcolsep}{4pt} % Adjust space between columns
\renewcommand{\arraystretch}{1.2} % Adjust space between rows
\begin{tabular}{|p{3cm}|p{3cm}|p{4cm}|p{3cm}|p{3cm}|}
\hline
\textbf{Category} & \textbf{Name of Algorithm} & \textbf{Variants} & \textbf{Type} & \textbf{Implementations} \\ \hline

\multirow{2}{*}{Lattice-based} 
& CRYSTALS-Kyber & Kyber512, Kyber768, Kyber1024 & Key Encapsulation & liboqs\cite{liboqs}, CIRCL\cite{circl}, pqm4\cite{pqm4}, AWS-LC \cite{aws-lc} \\ \cline{2-5}
& FRODOKem\cite{frodo2016} & FRODO-640, FRODO-976, FRODO-1344 & Key Encapsulation & liboqs, CIRCL \\ \cline{2-5}
&SABER &  LightSABER, SABER, FireSABER & Key Encapsulation & N/A \\ \cline{2-5}
& NTRU & NTRUEncrypt, NTRU-HRSS-KEM, and NTRU Prime & Key Encapsulation & liboqs, CIRCL \\ \cline{2-5}
& CRYSTALS-Dilithium & Dilithium-2, Dilithium-3, Dilithium-5 & Digital Signature & liboqs, CIRCL, pqm4 \\ \cline{2-5}
& FALCON 

% \cite{FALCON2018} 

& FALCON-512, FALCON-1024 & Digital Signature & liboqs, pqm4\\ \cline{2-5}

& ML-DSA\cite{ml-dsa} & ML-DSA-44, ML-DSA-65, ML-DSA-87 & Digital Signature & liboqs\\ \hline

\multirow{2}{*}{Code-based} 
& Classic McEliece & Classic-McEliece-348864, Classic-McEliece-460896, Classic-McEliece-6688128, Classic-McEliece-6960119, Classic-McEliece-8192128 & Key Encapsulation & liboqs\\ \cline{2-5}
& BIKE \cite{bike2021} & BIKE-L1, BIKE-L3, BIKE-L5 & Key Encapsulation & liboqs, pqm4 \\ \cline{2-5}
& HQC\cite{hqc2018}  & HQC-128, HQC-192, HQC-256 & Key Encapsulation & liboqs \\ \hline

\multirow{2}{*}{Hash-based} 
& SPHINCS+ \cite{sphincs2019} & SPHINCS+-SHA2-128-simple,
SPHINCS+-SHA2-192-simple, SPHINCS+-
SHA2-256-simple & Digital Signature & liboqs \\ \cline{2-5}
&XMSS\cite{xmss2011} & XMSS, XMSS-MT & Digital Signature & N/A \\ \hline

\multirow{2}{*}{Multivariate polynomial based} 
& Rainbow & Rainbow-I, Rainbow-III, Rainbow-V & Digital Signature & N/A \\ \cline{2-5}
& GeMSS & GeMSS128, GeMSS192, GeMSS256 & Digital Signature & N/A\\ \hline

\multirow{1}{*}{Isogeny-based} 
& SIKE & SIKE-p434, SIKE-p503, SIKE-p610 & Key Encapsulation & liboqs\\ \hline

\multirow{2}{*}{Zero Knowledge-based}
& PICNIC\cite{picnic2019} & PICNIC2-L1-FS, PICNIC2-L3-FS, PICNIC2-L5-FS & Digital Signature & N/A \\ \hline

\end{tabular}
\caption{Post-Quantum Cryptography Algorithms}
\label{tab:postquantum}
\end{table*}

\section{Protocol Analyses}

In this section, we examine nine protocols that collectively secure communication across the Internet stack ---from transport-layer encryption (TLS, QUIC) and network-layer tunnelling (IPsec, OpenVPN) to infrastructure services (BGP, DNSSEC), remote administration (SSH), federated identity (OpenID Connect), and end-to-end messaging (Signal Protocol). For each protocol, we describe its current cryptographic foundations, assess its vulnerability to quantum attack, and review the post-quantum migration efforts undertaken to date, including standardisation progress, experimental deployments, and open challenges. The protocols are ordered to reflect their dependencies: TLS is presented first as several subsequent protocols ---QUIC, OpenVPN, and OpenID Connect ---build directly upon its handshake.

\subsection{TLS (Transport Layer Security)}
TLS is the dominant protocol for secure communication on the Internet, used to encrypt web (HTTPS) and application traffic. It provides an authenticated key exchange (the TLS handshake) to establish shared secrets, and uses those secrets for symmetric encryption of application data.

\textbf{Current pre-quantum cryptography. } TLS 1.3, the latest version of the protocol, mandates ephemeral key exchange for every session. By default, this uses elliptic-curve Diffie–Hellman (ECDHE) over X25519 or NIST P-256, ensuring that each connection derives fresh keying material. Older versions such as TLS 1.2 also supported finite-field DH and static RSA key transport, but these modes are now deprecated in favour of the forward-secrecy guarantees provided by ephemeral exchange. Server authentication --- and optionally client authentication --- is performed via X.509 certificates carrying RSA or ECDSA public keys. During the handshake, the server signs a transcript of the exchange with its private key to prove possession, and the client validates this signature against a trusted certificate chain. In practice, RSA-2048 certificates remain widespread, though ECDSA P-256 certificates are increasingly common due to their smaller size and faster verification. Once the handshake completes, all application data is protected using symmetric AEAD ciphers --- typically AES-128-GCM, AES-256-GCM, or ChaCha20-Poly1305 --- which provide both confidentiality and integrity in a single construction. These symmetric algorithms are considered safe against quantum attack, as Grover's algorithm offers only a quadratic speedup that is not practical at scale with current key sizes.

\textbf{Quantum safety. } Not quantum-safe currently. The ECDHE key exchange that underpins every TLS 1.3 session would be broken by Shor's algorithm, allowing an eavesdropper who captures the handshake to recover the shared secret and decrypt all application data. TLS 1.3's mandatory use of ephemeral key exchange provides forward secrecy --- once session keys are discarded, past sessions cannot be retroactively compromised by breaking the server's long-term key alone --- but this protection is precisely what the "harvest-now, decrypt-later" threat circumvents: an adversary recording today's handshakes can derive session keys once a sufficiently powerful quantum computer becomes available. Server authentication is equally at risk, as both RSA and ECDSA signatures can be forged by a quantum adversary, enabling impersonation of any server whose certificate chain relies on these schemes. Since the entire Web PKI --- from root CAs down to end-entity certificates --- is built on RSA and ECDSA, this vulnerability is systemic rather than confined to individual deployments. The symmetric AEAD ciphers used after the handshake (AES-GCM, ChaCha20-Poly1305) remain quantum-safe, as does the HKDF-based key derivation. In summary, TLS 1.3's quantum exposure lies entirely in the handshake: key exchange and certificate-based authentication are vulnerable, while the symmetric data protection remains sound. Opus 4.6Extended

\subsubsection{PQC Migration Efforts.} 
\textit{Hybrid and Post-Quantum Key Exchange:} Several hybrid key exchange schemes for TLS 1.3 have been tested. For example, Google and Cloudflare's CECPQ experiments combined classical ECDH with post-quantum KEMs (e.g.\ X25519 + NTRU or SIKE in CECPQ2). Building on CECPQ2, Bernstein et al.~\cite{bernstein2022opensslntru} introduce a batch key generation technique for \texttt{sntrup761} that outperforms both \texttt{ntruhrss701} and pre-quantum schemes (NIST P-256, X25519) in TLS sessions per second, while requiring only minimal changes to OpenSSL and no application-level modifications. More recently, industry tests have focused on the NIST-selected KEM CRYSTALS-Kyber. A preliminary variant of Kyber has already been deployed in TLS 1.3 by Google Chrome and Cloudflare in hybrid mode to counter the ``harvest-now, decrypt-later'' threat. In fact, as of early 2024, ~2\% of Cloudflare's TLS 1.3 connections are using a post-quantum key agreement, a number expected to reach double digits by end of 2024. To systematically assess the algorithm-level trade-offs involved in such deployments, Paquin et al.~\cite{paquin2020benchmarking} measure TLS~1.3 handshake times for a wide range of NIST candidate KEMs and signature schemes using the Open Quantum Safe (OQS) fork of OpenSSL under emulated network conditions with varying latency, bandwidth, and packet loss. The broader engineering challenges of integrating PQC into TLS and SSH---including algorithm negotiation and key combination for hybrid modes---are addressed by Crockett et al.~\cite{crockett2019prototyping}, who report on prototype implementations in Amazon s2n, OQS-OpenSSL, and OQS-OpenSSH that served as foundational infrastructure for much of the subsequent PQ-TLS research.

\textit{Standards Progress:} The IETF TLS Working Group is actively standardizing PQC for TLS 1.3. One draft defines new TLS Named Groups for ML-KEM (Kyber) at various security levels (512, 768, 1024 bits) to enable pure post-quantum key agreement in TLS~\cite{ietf-draft-mlkem}. Another draft describes a hybrid key exchange design to combine classical ECDHE with a PQ KEM in the TLS handshake~\cite{ietf-draft-hybrid-tls}. These drafts will allow TLS peers to negotiate hybrid or PQ key exchanges in a standard way (several test implementations exist using OpenSSL with liboqs). The most comprehensive survey of the resulting landscape is provided by Alnahawi et al.~\cite{alnahawi2024comprehensive}, who classify existing post-quantum TLS proposals into three main categories, conduct unified performance simulations, and identify open research problems. Their benchmarks confirm that hybrid Kyber+X25519 key exchange adds only modest overhead, while more conservative schemes like FrodoKEM remain practical albeit slower.

\textit{Post-Quantum Certificates:} Migrating TLS authentication certificates is more challenging. There are ongoing efforts to define PQC signature algorithms in X.509 and TLS. IETF has proposals for composite/hybrid certificates that include both a classical and a PQ signature/public-key, so that browsers can continue to accept classical signatures while gradually adding trust in PQ signatures \cite{ietf-lamps-pq-composite-sigs-15}. Until standard PQ certificates are widely supported, any deployment of PQ signatures is ad-hoc (for instance, a server could offer a parallel PQ certificate chain in addition to the normal one)~\cite{cloudflare2024pq}. In a companion pair of studies, Sikeridis et al.~\cite{sikeridis2020assessing,sikeridis2020post} evaluate the joint overhead of post-quantum key exchange and authentication in TLS~1.3 and SSH, reporting latency increases of 1--300\% for TLS depending on the algorithm combination, and propose mixing different PQ signature algorithms across the certificate chain to achieve better latency trade-offs than a uniform scheme. A fundamentally different approach to the authentication overhead is taken by Schwabe et al.~\cite{schwabe2020post}, who propose KEMTLS, an alternative TLS~1.3 handshake that replaces signatures entirely with KEM-based authentication. Since post-quantum KEMs are generally more compact and faster than post-quantum signatures, KEMTLS significantly reduces handshake data volume and is formally proven secure in the standard model, though it delays authenticated server application data by one additional round trip.

\textit{Performance and Feasibility:} On dedicated hardware, Sosnowski et al.~\cite{sosnowski2023performance} conduct both black-box and white-box measurements of PQC in TLS~1.3, finding that most PQC algorithms are competitive with---or faster than---traditional schemes, that hybrid algorithms introduce negligible overhead, and that Dilithium outperforms RSA-2048 at all security levels. They also highlight that large PQC key sizes can trigger additional round trips in constrained network environments. For embedded platforms, B\"{u}rstinghaus-Steinbach et al.~\cite{burstinghaus2020post} are the first to integrate PQC into TLS, adding Kyber and SPHINCS+ to the mbed~TLS library and benchmarking across four ARM and Xtensa-based boards; they find that Kyber performs comparably to ECC, while SPHINCS+ poses challenges primarily on the server side. In the IoT domain specifically, Gonzalez and Wiggers~\cite{gonzalez2022kemtls} compare KEMTLS to post-quantum TLS~1.3 on a Cortex-M4 platform using WolfSSL across broadband, LTE-M, and Narrowband IoT scenarios, showing that KEMTLS reduces handshake time by up to 38\%, lowers peak memory consumption, and saves traffic volume---underscoring the benefits of signature-free authentication for resource-constrained devices.

\textbf{Challenges. }Despite the progress outlined above, several challenges remain. Post-quantum public keys, ciphertexts, and signatures are substantially larger than their classical counterparts, often causing handshake messages to exceed typical MTU sizes and triggering fragmentation or additional round trips. This size inflation is especially problematic for authentication, where each NIST signature finalist presents awkward trade-offs for certificate migration --- large outputs for Dilithium and SPHINCS+, and hardware-specific requirements for Falcon --- which partly explains why post-quantum key exchange has outpaced authentication in real-world deployment. Proposals such as KEMTLS and mixed certificate chains offer partial relief but require non-standard changes to the TLS state machine or certificate ecosystem, and the problem is compounded on embedded and IoT devices where limited RAM, bandwidth, and computational power amplify the cost of larger cryptographic objects.
Beyond performance, most proposals that modify the TLS handshake flow lack rigorous formal security proofs, making it difficult to confirm that protocol-level changes preserve the guarantees TLS 1.3 was designed to provide. Deployment must also contend with a heterogeneous ecosystem of clients, servers, middleboxes, and certificate authorities, where hybrid modes increase negotiation complexity and composite certificates raise backward-compatibility questions that remain unresolved.

\subsection{SSH}
SSH is a protocol for secure remote login and other secure network services. It provides an encrypted channel between a client and server, typically used for administration of systems (e.g. ssh into a Linux server). It uses its own key exchange and authentication mechanisms at the application layer (not TLS). 

\textbf{Current cryptography.}
The SSH handshake supports Diffie-Hellman key exchange. Modern SSH (e.g. OpenSSH) by default uses elliptic-curve DH over Curve25519 (sometimes called X25519) for key exchange. It also supports classic DH groups (modp) and recently even hybrid methods (see below). In older or alternative configs, an RSA-based key exchange is possible (though rarely used now).

The server proves its identity by possessing a host key. Common host key types are RSA, ECDSA, or Ed25519 (EdDSA). OpenSSH, for instance, now defaults to an Ed25519 host key because of its strong security and small size, though RSA 3072+ or ECDSA P-256 keys are also seen. The host key is used to sign the key exchange to authenticate the server to the client.

After the secure channel is established, the client can authenticate to the server either with a password or using SSH public key authentication. The latter typically involves the client’s RSA, ECDSA, or Ed25519 key pair. This step is separate from the key exchange.

SSH then derives symmetric session keys (usually AES-256 or ChaCha20 for encryption, and HMAC-SHA2 or similar for integrity). These symmetric algorithms are quantum-safe (AES and HMAC are safe if key sizes/hashes are large enough).

\textbf{Quantum safety.} Not quantum-safe currently. The X25519 elliptic curve Diffie–Hellman that secures most SSH sessions today would be broken by a quantum computer, just like in TLS. An eavesdropper recording an SSH session’s initial key exchange could later derive the session keys and decrypt the entire session if they have a QC. (SSH, like TLS, provides forward secrecy—once the session is over and keys discarded, only a captured transcript plus broken DH would yield plaintext.). RSA and ECDSA host keys can be forged with quantum computing, meaning an attacker could potentially impersonate a server if they can break the signature when the server authenticates in the handshake. However, note that SSH host keys are often verified out-of-band (through a known hosts file or TOFU model), not a CA, but the fundamental vulnerability remains if an attacker can forge the handshake signature, they can pose as the server. Ed25519 (being an elliptic curve scheme) is equally vulnerable to Shor’s algorithm. Symmetric encryption remains fine (e.g. AES-256 in SSH is okay against quantum, aside from requiring at most a doubling of key length in theory).

\textbf{PQC migration efforts.} Already in progress in implementations.

\textit{Hybrid key exchange in OpenSSH:} OpenSSH introduced a new key exchange method called sntrup761x25519-sha512@openssh.com, which is a hybrid of X25519 (ECDH) and a post-quantum KEM Streamlined NTRU Prime (sntrup761). This means the client and server perform both an X25519 exchange and an NTRU Prime exchange, and combine the results. An attacker would need to break both the classical and the post-quantum parts to recover the shared key. This was an early move to give SSH quantum resistance for the session key. As of OpenSSH 8.5, this hybrid KEX is supported (and was the default for a time). It has since been adjusted as the underlying NTRU Prime parameters evolved, but OpenSSH remains one of the first widely-used tools to include PQC.

\textit{Standardization:} Inspired by OpenSSH’s move, an Internet-Draft has been written to codify the X25519+sntrup761 hybrid key exchange for SSH in a standardized way \cite{ietf-draft-ntruprime-ssh}. This draft would make it easier for other SSH implementations to adopt the same method and ensure interoperability. So far, OpenSSH leads the charge, but we expect others (like LibreSSH, or SSH libraries) to follow when standards solidify.

Prototyping efforts by Crockett et al. \cite{crockett2019prototyping} demonstrated that SSH can practically integrate hybrid (classical+PQ) key exchanges and authentication, but exposed real implementation constraints such as message size limits and negotiation complexity. Measurement studies by Sikeridis et al \cite{sikeridis2020assessing} showed that in practice, handshake overhead is driven more by signature and certificate size—and even TCP congestion behavior—than by raw KEM computation, indicating that network-layer dynamics are central to deployability. Formal analyses by Duong et al. \cite{tran2023formal} revealed subtle authentication flaws in draft hybrid SSH designs and proposed verified fixes, while computational proofs in the PQ setting by Blanchet \& Jacomme \cite{blanchet2024post}; Bencina et al. \cite{benvcina2025post} provided stronger security models capturing ``harvest-now, decrypt-later'' threats and clarified that hybrid SSH can achieve ACCE-style guarantees under carefully defined assumptions. More recent protocol proposals by Qi \& Chen \cite{qi2025hpqke} push further by designing SSH-specific hybrid key exchanges (e.g., ECDH+CSIDH) that embed authentication into MAC-based confirmation rather than relying solely on PQ signatures.

\textbf{Challenges. } Post-quantum migration of SSH is not a simple primitive swap but a systemic redesign challenge. Hybrid key exchange must securely combine classical and PQ components without inducing downgrade or authentication flaws, while formal guarantees must now hold against quantum-capable adversaries under stronger models. At the same time, PQ signatures and larger key material significantly increase handshake size, making network effects (TCP congestion, fragmentation, message limits) a primary performance bottleneck rather than raw computation. Compounding this, SSH must preserve backward compatibility during a prolonged transition, manage expanded negotiation complexity, and operate amid evolving standardization and primitive maturity.

\subsection{QUIC}
QUIC is a modern transport protocol originally developed by Google and now standardized by the IETF. It runs over UDP and provides stream multiplexing, low latency connection establishment, and integrated security equivalent to TLS 1.3. HTTP/3 is built on top of QUIC. Essentially, QUIC implements a TLS 1.3 handshake within its own protocol to set up encryption between client and server, then carries data streams.

\textbf{Current cryptography.}
QUIC uses the TLS 1.3 handshake for cryptographic negotiation and key exchange, but encapsulates it in QUIC packets. So it uses the same algorithms as TLS 1.3, typically an ECDHE (X25519) key exchange, with server authentication via an RSA/ECDSA certificate, and symmetric AES-GCM or ChaCha20 encryption for packets.
After the handshake, QUIC encrypts all payload and most header fields using the derived TLS keys. Integrity is ensured via AEAD (the GCM or Poly1305 tags).
In short, QUIC’s security is equivalent to TLS 1.3’s security. The main difference is QUIC has some protocol-level integrity for its header (using a header protection algorithm), but that also relies on symmetric crypto (e.g. an AES or ChaCha mask) derived from the handshake keys.

\textbf{Quantum safety.} Not quantum-safe currently. Since QUIC relies entirely on the TLS 1.3 handshake for cryptographic negotiation, it inherits the same quantum vulnerabilities. The X25519 key exchange underpinning most QUIC connections would be broken by Shor's algorithm, allowing an eavesdropper who records the handshake to later derive session keys and decrypt all traffic --- including the payload and header fields that QUIC encrypts beyond what TLS traditionally protects. Server authentication relies on the same RSA or ECDSA certificates as TLS 1.3, both vulnerable to quantum forgery. QUIC's header protection mechanism uses symmetric primitives (AES or ChaCha20) and is quantum-safe in itself, but only if the handshake keys from which it derives were securely established --- which they would not be under a quantum adversary. The symmetric AEAD encryption (AES-GCM or ChaCha20-Poly1305) used for packet protection remains sound against quantum computers. In short, QUIC's quantum vulnerability profile is effectively identical to that of TLS 1.3: key exchange and authentication are at risk, while symmetric encryption and integrity mechanisms remain safe.

\textbf{PQC migration efforts.} Cloud-scale experiments by Raavi et al.\cite{raavi2022quic, raavi2023post} show that QUIC generally outperforms TCP/TLS even under PQ authentication, maintaining lower handshake latency and variance across global RTT conditions, with lattice-based signatures (Dilithium, Falcon) proving practical while larger schemes increase overhead. Cryptography-centric dissection by Kempf at al.\cite{kempf2024quantum} reveals that handshake byte size—not raw computation—is the dominant cost driver, making compact KEMs like Kyber and efficient signatures like Dilithium or Falcon viable, while SPHINCS+ imposes prohibitive latency due to large signatures. Comprehensive end-to-end evaluations by Rigon at al. and Montenegro et al. \cite{rigon2025comprehensive, montenegro2025comparative} confirm that ML-KEM/Kyber consistently delivers the best tradeoff across handshake latency, throughput, CPU, and memory, that hybrid ML-KEM+ECDHE adds only marginal overhead while strengthening transitional security, and that QUIC dampens PQ penalties better than TLS, especially under lossy conditions. Embedded-system analysis by Dong et al.\cite{dong2025epquic} further demonstrates that Kyber can outperform classical ECDH even on resource-constrained ARM platforms and that QUIC’s fast UDP-based handshake helps offset PQ computational costs, though high-security levels and hash-based signatures significantly inflate latency.

\textbf{Challenges. }The migration of QUIC to post-quantum cryptography benefits from the protocol's UDP-based design and reduced round-trip handshake, which naturally dampens some of the latency penalties introduced by larger PQ primitives. However, several challenges remain. The dominant cost driver in post-quantum QUIC is handshake byte size rather than raw computation, meaning that algorithms with large public keys or signatures --- particularly SPHINCS+ --- can impose prohibitive overhead even when their computational cost is manageable. Since QUIC multiplexes all handshake and application data over UDP without the segmentation guarantees of TCP, oversized PQ handshake messages risk exceeding path MTU limits and triggering fragmentation at the IP layer, which interacts poorly with middleboxes and firewalls that may drop or reorder UDP fragments. While compact KEMs like Kyber and efficient signatures like Dilithium or Falcon have proven practical, the lack of a standardised post-quantum QUIC profile means that algorithm selection and hybrid negotiation remain ad hoc across implementations. On resource-constrained and embedded platforms, QUIC's fast handshake helps offset PQ computational costs at lower security levels, but high-security parameter sets and hash-based signatures can still inflate latency to the point where the protocol's speed advantages over TLS are eroded. Finally, QUIC's tight integration of transport and cryptography --- while architecturally elegant --- means that any PQ-related changes to the handshake have implications for congestion control, connection migration, and 0-RTT resumption, interactions that are not yet well understood and have received limited formal analysis.

\subsection{IPsec (IKEv2)}
IPsec is a suite of protocols for securing IP communications (VPNs) at the network layer. It consists of the Internet Key Exchange (IKEv2) protocol for negotiating cryptographic keys and security parameters, and the AH/ESP protocols for authenticating and encrypting IP packets.

\textbf{Current Cryptography.} Current cryptography. IPsec operates in two phases. The first phase uses IKEv2 to negotiate security parameters and establish a shared secret. IKEv2 performs a Diffie–Hellman key exchange — typically using finite-field DH groups (modp2048, modp3072) or elliptic-curve DH (e.g. Curve25519 or NIST P-256) — to derive keying material. Peers authenticate each other either through digital signatures (RSA or ECDSA backed by X.509 certificates), which is common in site-to-site VPN deployments, or through pre-shared symmetric keys, which are simpler but less scalable. The second phase establishes one or more IPsec Security Associations (SAs) that protect the actual data traffic. The Encapsulating Security Payload (ESP) protocol handles packet-level protection, using symmetric ciphers such as AES-CBC or AES-GCM for encryption and HMAC-SHA-256 or the GCM authentication tag for integrity. These symmetric algorithms are generally considered quantum-resistant provided sufficiently large keys are used. IKEv2 also supports periodic re-keying to refresh session keys over long-lived tunnels, performing a new DH exchange each time to maintain forward secrecy.

\textbf{Quantum Safety.} Not quantum-safe currently. The Diffie–Hellman key exchange at the heart of IKEv2 --- whether finite-field or elliptic-curve --- would be broken by Shor's algorithm, allowing an eavesdropper who records the IKE handshake to later recover the shared secret and decrypt all ESP-protected traffic in that session. Unlike TLS, where sessions are typically short-lived, IPsec tunnels often persist for extended periods with periodic re-keying; if the underlying DH exchange is compromised, all traffic within a tunnel's lifetime is exposed. Peer authentication via RSA or ECDSA signatures is equally vulnerable --- a quantum adversary capable of forging these signatures could impersonate a VPN gateway and establish rogue tunnels, a particularly severe risk in site-to-site deployments where IPsec protects entire network segments rather than individual connections. Pre-shared key authentication, while not directly broken by quantum computers, is only as strong as the key exchange it accompanies; if the DH-derived secret is compromised, the PSK alone does not protect session confidentiality. The symmetric encryption (AES) and integrity mechanisms (HMAC-SHA2) used by ESP after the handshake remain quantum-safe with current key sizes. In summary, IPsec's quantum vulnerability centres on IKEv2: both the key exchange and certificate-based authentication are at risk, while the symmetric data-plane protections remain sound.

\textbf{PQC Migration of IPsec.} \newline
\textit{Hybrid Key Exchange:} Recent standards allow mixing a pre-shared key (PSK) into the IKEv2 key derivation, yielding quantum-resistant keys even if the DH exchange is later broken. RFC 8784 (2020) introduces this ``Post-quantum Preshared Key'' extension to augment IKEv2 with entropy from a pre-shared secret~\cite{rfc8784}. Early practical exploration of this direction is provided by Herzinger et al.~\cite{herzinger2021real}, who examine hybrid key exchange in real-world IKEv2 deployments, noting that the lack of confidence in any single post-quantum algorithm motivates combining at least two schemes so that the shared secret remains secure as long as one prevails; however, the large payloads of some PQ algorithms require significant protocol-level changes. Blanco-Romero et al.~\cite{blanco2025hybrid} push this further by proposing a hybrid architecture for IPsec that combines classical, post-quantum, and quantum key distribution (QKD) sources into the key derivation process, providing defence in depth across multiple layers of cryptographic assurance.

\textit{Multiple Key Exchanges:} The new RFC 9370 (2023) extends IKEv2 to perform multiple key exchanges (e.g.\ an ECDH plus a post-quantum KEM) during session setup. This allows negotiating one or more PQC algorithms alongside the classical DH. The shared key can be formed by combining secrets such that an attacker must break all key exchanges to defeat the security. In other words, if at least one component algorithm is quantum-resistant, the final IKEv2 shared secret is quantum-safe~\cite{rfc9370}. A detailed performance breakdown under this model is provided by Bae et al.~\cite{bae2022performance}, who evaluate a range of NIST Round~3 KEMs (Kyber, NTRU, Saber) and regionally developed algorithms within the strongSwan IPsec implementation via liboqs, analysing execution speed and packet size across security levels. Their results show that higher security levels generally increase latency and packet sizes, with Kyber offering the best balance between security and performance. Gazdag et al.~\cite{gazdag2023quantum} take these efforts towards standardisation by executing the first steps needed for quantum-resistant VPNs on both Layer~2 (MACsec/MKA) and Layer~3 (IPsec/IKEv2), identifying the necessary protocol modifications and testing them in practice.

\textit{Deployment and Testing:} Despite these standardisation efforts, Twardokus et al.~\cite{10.1145/3733825.3765281} demonstrate that the proposed IETF RFCs for quantum-resistant IKEv2 remain largely untested under realistic conditions. Using a reproducible testbed deployed over both lossy wireless links and the internationally distributed FABRIC testbed, they reveal severe bottlenecks under high round-trip times and non-trivial packet loss, reporting a 400--1000-fold increase in data overhead over high-loss wireless links. On the physical layer side, Lawo et al.~\cite{lawo2024wireless} report on the first experimental IPsec tunnel secured by Falcon, Dilithium, and Kyber over both wireless and fiber-optic links; since strongSwan did not natively support PQC at the time, they perform authentication and key exchange externally and establish the IPsec connection using the exchanged pre-shared key.

\textbf{Challenges. } The migration of IPsec to post-quantum cryptography faces challenges that mirror --- and in some cases exceed --- those encountered in TLS. IKEv2 was designed around the assumption of compact key exchange payloads, and its fragmentation mechanisms are less mature than those in TLS 1.3, making the handshake particularly sensitive to the message size inflation introduced by PQ algorithms. In practice, this can translate to orders-of-magnitude increases in data overhead over lossy or bandwidth-constrained links, suggesting that VPN deployments over wireless, satellite, or high-latency networks may require protocol redesign rather than simple algorithm substitution. Unlike TLS, where hybrid key exchange has already seen large-scale deployment by browser vendors and CDNs, IPsec adoption of PQC remains largely confined to research prototypes and testbeds, with no major commercial VPN product shipping post-quantum IKEv2 by default. The lack of native PQC support in widely used IPsec daemons further slows adoption, as workarounds like performing key exchange externally and injecting pre-shared keys add operational complexity and preclude standard re-keying. Finally, formal security analysis of post-quantum IKEv2 extensions is still in its early stages, with only limited automated proofs available for the modified handshake flows --- a gap that needs to be closed before these extensions can be confidently standardised and deployed at scale.

\subsection{OpenVPN}
OpenVPN is a popular open-source VPN protocol/software that secures IP traffic at the transport layer (it can be thought of as similar in purpose to IPsec, but in user space). It uses TLS (or a TLS-like handshake) to establish a secure tunnel between client and server, and then encrypts IP packets or TCP streams through that tunnel.

\textbf{Current cryptography.} OpenVPN uses a TLS handshake --- typically TLS 1.2, though newer versions support TLS 1.3 --- to establish a secure control channel between the VPN client and server. During this handshake, peers authenticate using X.509 certificates, most commonly RSA-2048 or RSA-4096 for the server, with clients presenting either their own certificate or a pre-shared key depending on the deployment configuration. Key exchange follows the standard TLS model, using ECDHE (e.g. X25519 or NIST P-256) or, in older configurations, RSA key transport to derive session keys. In essence, the authentication and key exchange algorithms are whatever the underlying TLS profile dictates, making OpenVPN's handshake security equivalent to that of web TLS. Once the handshake completes, OpenVPN encrypts user traffic on a separate data channel using symmetric ciphers --- most commonly AES-256 in CBC or GCM mode, or ChaCha20-Poly1305 for better performance on devices lacking hardware AES acceleration --- with HMAC-SHA-256 providing integrity where AEAD is not used. A separate control channel handles session management, re-keying, and keepalive messages, and is protected directly by the TLS-derived keys. OpenVPN also supports an optional static pre-shared key mode (tls-auth or tls-crypt) that wraps the control channel in an additional symmetric encryption layer, providing a first line of defence against denial-of-service attacks and unauthenticated probing before the TLS handshake even begins.

\textbf{Quantum safety.} Not quantum-safe currently, and inherits all of TLS's quantum vulnerabilities since its security model is built directly on top of a TLS handshake. The ECDHE or RSA key exchange used during session establishment would be broken by Shor's algorithm, allowing an adversary who records the VPN tunnel setup to later derive session keys and decrypt all encapsulated user traffic --- effectively stripping away the VPN's confidentiality entirely. The RSA or ECDSA certificates used for peer authentication are equally vulnerable to quantum forgery, meaning an attacker could impersonate a VPN server or client and establish rogue tunnels. This is particularly concerning in enterprise and remote-access scenarios where OpenVPN protects sensitive internal network traffic, as a single compromised handshake exposes not just one connection but all traffic routed through that tunnel. The tls-auth and tls-crypt pre-shared key layers, being symmetric, are quantum-safe and would continue to prevent unauthenticated access to the control channel, but they do not protect the key exchange itself --- they merely gate access to it. The symmetric data-channel encryption (AES-256, ChaCha20) and integrity mechanisms (HMAC-SHA-256, GCM tags) remain quantum-safe. In summary, OpenVPN's quantum exposure is identical to that of the TLS version it runs on: key exchange and certificate authentication are at risk, while symmetric protections remain sound, and the protocol's VPN-specific additions (tls-auth, tls-crypt) offer no mitigation against a quantum adversary targeting the handshake.

\textbf{PQC migration efforts.} 
\textit{TLS 1.3 in OpenVPN:} Since OpenVPN can now use TLS 1.3 (which has a standard extension mechanism for new key exchange groups), once TLS 1.3 gets official PQ groups (e.g. via the draft for ML-KEM Kyber), OpenVPN will be able to use them out-of-the-box by relying on the TLS library.

\textit{Microsoft Research’s PQCrypto-VPN:} Microsoft published an experimental fork of OpenVPN that integrates post-quantum algorithms \cite{pqvpn-microsoft}. This project used the Open Quantum Safe (liboqs) library to add PQ key exchange into OpenVPN’s TLS handshake. It allowed testing of algorithms like FrodoKEM or others in a VPN scenario, evaluating performance and interoperability. The fork demonstrated that one can have a working OpenVPN with quantum-resistant key exchange, but it was for research (not production).

\textit{OpenVPN community developments:} OpenVPN’s developers have been aware of PQC. A 2022 patch on the OpenVPN mailing list enabled support for quantum-safe and hybrid key exchanges via OpenSSL 3.0’s provider interface
\cite{openvpn-patch}. OpenVPN can be built with OpenSSL, and OpenSSL 3.0+ can dynamically load the Open Quantum Safe (OQS) provider, which includes PQ algorithms. The patch ensured that if OpenSSL offers a PQ KEM (like Kyber) as a TLS group, OpenVPN can negotiate it just like any other cipher suite. This patch indicates active interest and likely will be part of OpenVPN 2.6+ or 2.7.

\subsection{BGP and RPKI}
BGP is the routing protocol that connects the global Internet, used between autonomous systems (AS) to exchange network reachability information. By default, BGP has minimal security, which led to extensions like RPKI and BGPsec to provide origin authentication and path validation for routes.

\textbf{Current cryptography.} \textit{Classic BGP} uses TCP with MD5 or TCP-AO for authentication of BGP sessions. These are symmetric integrity checks (MD5 or HMAC) on each packet – while MD5 is outdated, TCP-AO with a strong HMAC (SHA-1/256) can protect BGP sessions from tampering. Symmetric keys here are not threatened by quantum attacks if chosen with sufficient length. \textit{The Resource Public Key Infrastructure (RPKI)} provides a way to verify that an AS is authorized to originate a route for a given IP prefix. RPKI uses X.509 certificates and digitally signed objects (ROAs). Current RPKI deployments typically use RSA or ECDSA for those signatures (e.g. a common profile is RSA-2048 or ECDSA P-256 for certificates and ROAs). \textit{BGPsec} is an extension where each AS cryptographically signs the routing announcements it propagates. BGPsec uses ECDSA P-256 with SHA-256 as its mandatory algorithm for signing BGP UPDATE messages. Each BGPsec speaker has a signing key (provisioned via RPKI certificates) to sign route updates, and receivers validate the signature chain.

\textbf{Quantum safety.} Not quantum-safe. BGPsec’s use of ECDSA means that a quantum attacker could forge route announcements or invalidate legitimate ones by breaking the signatures. The RPKI’s use of RSA/ECDSA for certificates and ROAs is similarly vulnerable – a quantum-capable adversary could potentially hijack IP prefixes by forging RPKI certificate material or BGPsec signatures. While the symmetric session protections (TCP-AO with HMAC) would remain secure, they do not protect against attacks on route authenticity. In essence, the integrity of routing information relies on digital signatures that are quantum-vulnerable. It’s worth noting that BGP as a whole has not fully deployed BGPsec globally (adoption is limited due to complexity), but to the extent we rely on RPKI and BGPsec for security, those aspects are at risk in a post-quantum scenario.

\textbf{PQC migration efforts.} Very early stage. There is currently no deployed post-quantum version of BGPsec or RPKI, but the need is recognized.

The BGPsec design anticipated the need for cryptographic agility. RFC 8608 explicitly notes that ``BGPsec will require adoption of updated key sizes and a different set of signature and hash algorithms over time'', and that the profile should be updated with new algorithms when appropriate \cite{rfc8608}. This provides a pathway for introducing PQC algorithms into BGPsec.

 Academic groups and NIST have started looking at PQ deployment for BGP – for example, NIST’s BGP Secure Routing Extension (BGP-SRx) test suite could be used to experiment with alternate algorithms \cite{nist-bgpsrx, nist-brite}.

 Doesburg \cite{doesburg2025post} argues that RPKI’s reliance on RSA makes it vulnerable to future quantum attacks and shows that post-quantum migration—particularly via hybrid signatures like Falcon—can be feasible if performance overhead is mitigated through optimizations and a more practical, mixed-tree deployment model rather than rigid top-down transitions. Miesch et al. \cite{miesch2025poster} complement this by demonstrating that RPKI’s lack of algorithm agility is the fundamental barrier to both PQ and classical upgrades, and that naïve migration strategies risk overwhelming bandwidth and validation capacity; they propose a dual-tree (legacy + mixed) approach to enable incremental, incentive-compatible deployment.

\textbf{Challenges. } The notable obstacles include performance and message size, a BGPsec update may be processed by many routers, so signature verification must be fast. PQC signatures like Dilithium or Falcon are computationally heavier (though Falcon is quite fast in verification) and larger in size. A BGPsec UPDATE currently carries an ECDSA signature of 64 bytes per AS hop; a Dilithium signature could be ~2–3KB, which, with many hops, might considerably bloat BGP messages and strain router memory/CPU. Techniques to mitigate this (e.g. signing only critical parts, or using more compact PQ signatures if available) would be explored.
Another avenue is stateful hash-based signatures (like XMSS) for BGP, since routing updates are sequential and could use stateful signatures. These are quantum-safe and could potentially be efficient per update, but operational complexity (managing one-time keys across reboots/route changes) is a barrier.

\subsection{DNSSEC}
 DNSSEC adds cryptographic authenticity to DNS. It uses digital signatures to ensure that DNS responses (like the IP address for a domain name) haven’t been tampered with. It introduces a hierarchy of public keys: the DNS root, top-level domains, and individual domains each have key pairs used to sign DNS records. Resolvers (DNS clients) verify these signatures to trust the DNS answers.

At each zone (e.g. example.com), the zone signing key (ZSK) signs DNS record sets (RRsets). Those signatures (RRSIG records) are distributed via DNS. A chain of trust is formed: the parent zone signs a delegation to the child’s key (DS record). Ultimately, a validating resolver uses the root’s public key (the anchor) to verify the whole chain of signatures.

\textbf{Current cryptography.} DNSSEC supports various algorithms. The most widely used are RSA (with SHA-256, designated as algorithm 8 or 8+SHA-256 in DNSSEC), ECDSA (P-256 with SHA-256, and P-384 alg 14), and increasingly EdDSA (Ed25519 and Ed448). For example, the DNS root is currently signed with an RSA-2048 key (SHA-256 digest), and many TLDs and domains use either RSA-2048 or ECDSA P-256 for zone signing. Ed25519 is valued for its small signature size and is seeing adoption for smaller zones.
RSA keys in DNSSEC are often 2048-bit (some larger like 3072 or 4096 for root KSKs). ECDSA P-256 keys are 256-bit but provide similar security with much smaller signatures (64 bytes vs ~256 bytes for RSA-2048 sigs). Ed25519 likewise has 64-byte signatures.

\textbf{Quantum safety.} Not quantum-safe. All the algorithms in practical use for DNSSEC (RSA, ECDSA, EdDSA) are based on factorization or discrete log problems, which are breakable by a quantum computer. An attacker with a quantum computer could forge DNSSEC signatures, allowing them to create fake DNS records (e.g. trick users into connecting to an attacker’s IP instead of a real site) even if DNSSEC is deployed. They could impersonate entire zones by forging RRSIGs on DNS answers. They could also forge the chain of trust (e.g. pretend to be a parent TLD or even the root by forging signatures with the root’s private key).
The integrity of DNSSEC-signed data would collapse under a quantum adversary, negating the security DNSSEC provides (which is to prevent DNS spoofing/cache poisoning). DNSSEC does not protect confidentiality (DNS queries/responses are public), so the concern is integrity/authenticity. Quantum breaks would let an attacker undetectably sabotage name resolution.

\textbf{PQC migration efforts.} 
Industry and academic efforts have begun testing PQC within the DNS ecosystem. In 2023–2024, a collaboration between deSEC, SandboxAQ, and PowerDNS developers deployed experimental DNSSEC-signed zones using post-quantum signature schemes including Falcon-512, Dilithium-2, SPHINCS+, and the stateful hash-based scheme XMSS~\cite{dnssec-field-study}. These were not purely simulated environments --- the zones were operationally signed and served to real resolvers, allowing the researchers to observe how existing DNSSEC validators handle the larger key and signature sizes introduced by PQC, and to measure the resulting impact on query latency and UDP/TCP fallback behaviour. Modified versions of BIND 9 and PowerDNS were used as the authoritative servers in these experiments.

Muller et al. \cite{muller2020retrofitting} systematically evaluate NIST PQ signature candidates against DNSSEC’s practical limits (notably the ~1232-byte UDP safety threshold), showing that only a few schemes (e.g., Falcon-512) are even close to deployable without protocol changes and arguing that DNSSEC cannot treat PQ as a drop-in cryptographic swap. Goertzen \& Stebila \cite{goertzen2023post} introduce ARRF (application-layer request-based fragmentation), where oversized DNSSEC responses are split and explicitly requested at the DNS layer rather than relying on IP fragmentation or TCP fallback; their implementation with Falcon, Dilithium, and SPHINCS+ shows lower resolution latency and, in some cases, lower bandwidth than TCP fallback, while identifying memory-exhaustion risks from adversarial fragment metadata. McGowan et al. \cite{mcgowan2025security} further analyze those ARRF security risks and propose mitigations to prevent resource-allocation attacks during reassembly. Across a sequence of works, Rawat and Jhanwar progressively explore increasingly radical design points for enabling post-quantum DNSSEC. They first propose QNAME-based fragmentation (QBF) \cite{rawat2023post}, which keeps the signature-based model intact but enables oversized PQ responses to be reconstructed over UDP in a single round trip using standard DNS records. They then optimize the fallback path with TurboDNS \cite{rawat2024post}, reducing the latency overhead of TCP fallback and incorporating cookie-based safeguards to mitigate abuse. Most recently, with SL-DNSSEC \cite{rawat2025quantum}, they move beyond fragmentation entirely, replacing signatures with a quantum-safe KEM+MAC construction to eliminate large signature overhead while preserving DNSSEC’s security goals. Pan et al. \cite{pan2024double} advocate double-signing (classical + PQ) combined with fragmentation as a transitional hedge against both quantum and immature-PQ risks. Finally, Schutijser et al. \cite{schutijser2025evaluating} provide an operator-focused evaluation, signing real TLD-scale zones (.nl, .se, .nu) with Falcon-512 and MAYO-2, showing PQ signing is operationally feasible but with measurable zone size and CPU trade-offs.

\textbf{Challenges.} The main obstacle to PQ adoption in DNSSEC is signature and key size inflation, which conflicts with DNS’s UDP-oriented design and practical $\sim$1232-byte response limit. Large PQ signatures frequently trigger IP fragmentation or TCP fallback, increasing latency, bandwidth use, and operational complexity, while also expanding zone sizes and raising signing and validation costs---especially for large operators.

Equally challenging is deployability at Internet scale. Fragmentation schemes, TCP optimizations, or dual-signing approaches introduce new protocol logic and potential attack surfaces (e.g., resource exhaustion), and must remain backward-compatible with heterogeneous resolvers and middleboxes. As a result, PQ migration in DNSSEC is fundamentally a systems and ecosystem coordination problem, not just a cryptographic upgrade.

\subsection{OpenID Connect}
OpenID Connect is an identity authentication protocol built on OAuth 2.0. It allows clients (like a web app) to verify a user’s identity by obtaining information (in an ID Token) from an Identity Provider (IdP) like Google, Microsoft, etc. For example, ``Log in with Google'' uses OpenID Connect. The ID Token is typically a JSON Web Token (JWT) signed by the IdP, which the client can cryptographically verify.

\textbf{Current cryptography.}
OIDC flows occur over HTTPS (HTTP + TLS) — for instance, the user is redirected to the IdP, and tokens are sent over HTTPS. So the transport security is TLS (with its classical crypto). Thus, all the TLS considerations (RSA/ECDHE, etc.) apply here as well.
The core of OIDC is the ID Token, which is a JWT (JSON Web Token). This is usually signed using JSON Web Signature (JWS). Common algorithms are RSA with SHA-256 (often called RS256 in JWT parlance) or ECDSA P-256 with SHA-256 (ES256). In practice, RS256 is very widely used as a default in many systems. 
The IdP has a key pair (RSA or EC) and publishes the public key via a JWKS (JSON Web Key Set) endpoint. The relying party (client) downloads that and uses it to verify the JWT signature, ensuring the token was issued by the genuine IdP.

OIDC can also encrypt tokens (using JSON Web Encryption – JWE) but this is less common; usually tokens are just signed and sent via HTTPS, relying on TLS for confidentiality.

Sometimes, after obtaining an ID Token, the client may call a UserInfo API at the IdP to fetch more profile info. That API call is over HTTPS and secured by OAuth access tokens, but not directly relevant to PQ crypto (mainly symmetric bearer token plus TLS).

\textbf{Quantum safety.} Not quantum-safe currently, though the threat model differs from transport-layer protocols. OIDC's transport security inherits all of TLS's quantum vulnerabilities — the X25519 key exchange and RSA/ECDSA certificates protecting HTTPS channels between the user, IdP, and relying party are all susceptible to Shor's algorithm. However, the more distinctive risk lies in the ID Token itself. The JWT signatures that form the trust anchor of OIDC are typically RS256 (RSA) or ES256 (ECDSA), both of which a quantum adversary could forge. An attacker able to break the IdP's signing key could mint arbitrary ID Tokens that any relying party would accept as genuine, effectively impersonating any user to any service that trusts that IdP --- a far broader blast radius than compromising a single transport session. Since IdP public keys are published via JWKS endpoints and often long-lived, a quantum adversary would not even need to intercept traffic; they could simply download the public key and compute the corresponding private key offline. Token encryption (JWE), where used, typically relies on RSA or ECDH key wrapping and would similarly be broken. The symmetric components --- such as AES content encryption within JWE or HMAC-based token validation in certain configurations --- remain quantum-safe. In summary, OIDC's quantum exposure is twofold: it inherits TLS's transport vulnerabilities, and it adds an application-layer risk through the forgery of JWT signatures that underpin the entire federated trust model.

\textbf{PQC migration efforts.} 
Schardong et al. \cite{schardong} present the first full-stack empirical evaluation of post-quantum OAuth 2.0 and OpenID Connect, replacing classical TLS key exchange and JWT signatures with Kyber and NIST finalist PQ signatures, and benchmarking a realistic multi-step OIDC deployment. They show PQ identity is technically feasible and competitive in low-latency settings, but increased key and signature sizes significantly amplify TLS handshake overhead under higher latency, making network effects --- not computation --- the primary bottleneck in practical PQ migration.

The IETF JSON Object Signing and Encryption (JOSE) working group and the COSE (CBOR Object Signing and Encryption) working group are standardizing how to use PQC algorithms in JSON/CWT tokens. Drafts exist for encoding CRYSTALS-Dilithium, Falcon, and SPHINCS+ signatures in JWT/COSE format \cite{ietf-draft-cose-pq}. 

\textbf{Challenges. } Post-quantum migration of OAuth 2.0 and OpenID Connect faces two core challenges: performance scaling and ecosystem coordination. Larger PQ keys, certificates, and signatures inflate TLS handshakes and token exchanges, compounding latency across the multi-step identity workflow and stressing bandwidth-constrained or mobile environments. At the same time, identity systems depend on tightly coupled components --- TLS stacks, JWT standards, discovery endpoints, certificate infrastructures, SDKs, and federated trust frameworks—so introducing PQ algorithms requires synchronized upgrades across the entire ecosystem. Hybrid fallback modes may ease transition but risk complexity and residual classical exposure, making PQ identity migration a systemic architectural shift rather than a simple algorithm swap.

\subsection{Signal Protocol}
The Signal Protocol (developed by Open Whisper Systems) is an end-to-end encryption protocol for secure messaging. It’s used in the Signal messenger, WhatsApp, and other apps to ensure that only the intended recipient of a message can read it. Signal Protocol introduced the Double Ratchet algorithm and X3DH (Extended Triple Diffie–Hellman) key agreement for asynchronous, secure key exchange between users. In brief, when two people start a conversation,
they perform an initial key agreement (X3DH) using a combination of long-term keys, semi-long-term (signed pre-keys), and one-time pre-keys – all based on Diffie–Hellman over elliptic curves. They then enter a Double Ratchet where each message uses a new symmetric key derived from the previous, providing forward secrecy and post-compromise security.

When User A wants to start a chat with User B, A obtains B’s identity key, B’s signed pre-key (and signature to verify it, using Ed25519), and one of B’s one-time keys. A then does DH between various combinations of these keys (A’s identity \& B’s pre-key, A’s ephemeral \& B’s identity, etc.) --- hence ``Triple DH'' --- to derive a shared master secret.

\textbf{Current cryptography.}
Signal uses Curve25519 (elliptic curve Diffie–Hellman on curve Montgomery 25519) for all its DH operations. Each user has an identity key pair (Curve25519), a signed pre-key (Curve25519, signed with the identity key via Ed25519 signature), and a set of one-time pre-keys (Curve25519). 

After X3DH, both parties have a shared secret and start exchanging messages. Each message uses a symmetric key evolved via the ratchet (which involves a DH each time someone sends a message to mix in new entropy, plus a hash ratchet). The ratchet DH again uses Curve25519: each party generates a new ephemeral Diffie–Hellman key for each ratchet step and exchanges public values, deriving new chain keys.

Signal protocol uses AES-256 (in CBC mode in older versions, now often XChaCha20 for the Double Ratchet in some implementations) and HMAC-SHA256 for encryption and authentication of messages. It uses HKDF (with SHA-256) for key derivation extensively.

There isn’t a global PKI; instead, users verify each other’s identity keys out-of-band (scanning QR codes or comparing fingerprints). This guards against MitM but relies on users to do verification. The identity keys are Ed25519 key pairs (since Ed25519 keys can be the same as Curve25519 keys through a mapping, Signal uses the Curve25519 key also as an Ed25519 signing key to sign the pre-key).

\textbf{Quantum safety.} Partially. Signal provides forward secrecy – once a message is delivered and keys ratchet forward, even if someone somehow got your current keys, they can’t decrypt old messages. However, against a quantum threat, the long-term and ephemeral keys are all ECC (Curve25519, Ed25519). A quantum attacker who records the initial handshake (the X3DH key exchange messages) could break the Curve25519 Diffie–Hellman and the Ed25519 signature (if trying active attack) and obtain the master secret that bootstraps the conversation. With that, they could decrypt all messages of that session (assuming they also recorded the ciphertexts). So the confidentiality of conversations is at risk if an adversary can harvest the key exchange data. This is analogous to TLS’s vulnerability.

Every message round uses Curve25519 DH for the ratchet step. Those DH exchanges are ephemeral and last only one step, but a quantum attacker who records them and later breaks them could link some part of the key evolution or possibly decrypt a particular message if they also had the relevant chain key state. However, because each DH is used to mix into a symmetric chain, breaking one ratchet step might only allow decrypting from that point onward (and not backward) – still a problem, but the design limits damage. Regardless, those DH steps are not quantum-safe either.

The message encryption (AES-256, HMAC-SHA256) is fine under quantum assumptions (AES-256 is strong, HMAC-SHA256 is okay, though SHA-256’s collision resistance is not critical here and Grover’s algorithm isn’t practical at scale).

If users verify fingerprints in person, that part is safe (it’s a human trust link). But if an attacker could forge an identity key signature, they might try to impersonate a user to someone who hasn’t verified (a quantum attacker could fake the Ed25519 signature on a malicious pre-key to perform a MitM if the users don’t verify identity fingerprints)

\textbf{PQC migration efforts:} In late 2023, the Signal team announced and started deploying an upgrade to the protocol called PQXDH (Post-Quantum Extended Diffie–Hellman). PQXDH augments the X3DH handshake by incorporating a post-quantum key agreement. Specifically, Signal added CRYSTALS-Kyber (Kyber1024) as a KEM alongside the existing X25519 exchange\cite{signal2024pqxdh}. In practice, when two Signal clients establish a session, they perform the normal elliptic-curve X3DH and additionally perform a Kyber encapsulation to the recipient’s static PQ public key. The two resulting shared secrets (one ECDH, one PQKEM) are combined (concatenated and hashed) to derive the session master key. This means an eavesdropper would need to break both the ECDH and the Kyber KEM to recover the key. Breaking just one (even the weak ECDH) is not enough.

PQXDH handshake is already implemented in the latest Signal clients (as of late 2023) and is being gradually enabled. Once both parties in a chat have updated to a PQXDH-capable version, new sessions they start will use the hybrid PQ handshake. After a sufficient upgrade period, Signal plans to require PQXDH for all new chats, fully deprecating the old X3DH. This is one of the first real-world deployments of post-quantum cryptography at the end-to-end application layer.

\textit{PQ for message signatures:} Signal doesn’t use message-level signatures (the protocol uses HMAC for authentication of messages, tied to the symmetric keys). So there isn’t an immediate need for a PQ digital signature in each message. The main signature use is the Ed25519 on the signed pre-key which establishes one’s identity key authenticity. That could be swapped out for a PQ signature (e.g. Dilithium) if needed in the future. But since that signature is only used to prevent an attacker from feeding a fake pre-key (a form of MitM), and since users are encouraged to verify keys out of band, the priority was clearly the confidentiality via PQXDH.

Early system-level analyses by Bobrysheva et al.\cite{bobrysheva2019post, bobrysheva2020post} established that Signal’s quantum vulnerability lies specifically in its Diffie–Hellman components --- X3DH initialization and the DH ratchet --- while its symmetric primitives (AES, HKDF) remain comparatively robust under Grover-adjusted assumptions. They further argued that messaging systems provide a practical testbed for PQ deployment and proposed replacing DH/X3DH with isogeny-based schemes such as SIKE, identifying integration issues around authentication and associated data binding. Moving from feasibility to cryptographic structure, Brendel et al.\cite{brendel2020towards} demonstrated that naively substituting DH with a post-quantum KEM breaks X3DH’s security guarantees because asynchronous prekey reuse is not naturally supported by standard KEM security models; they introduced the abstraction of split KEMs to capture the reuse and encapsulation patterns required by Signal. Building on this structural insight, Dobson and Galbraith\cite{dobson2022post} developed a formal Signal-specific AKE model and proposed SI-X3DH, a SIDH-based construction that preserves asynchronous key agreement while addressing adaptive static-key attacks. Finally, Brendel et al.\cite{brendel2022post} advanced the transition further by constructing a fully post-quantum asynchronous deniable key exchange (SPQR), combining KEM-based agreement with designated-verifier authentication to retain Signal’s core properties --- forward secrecy, break-in recovery, exposure resilience, and deniability --- under quantum-capable adversaries.

\textbf{Challenges. }The main challenges in this transition stem from the structural properties that make Signal powerful: asynchronous key establishment, semi-static prekeys, deniability, and strong exposure resilience. Most PQ KEMs are not naturally secure under key reuse patterns required by X3DH, and isogeny-based approaches introduce concerns about adaptive attacks and validation complexity. Preserving deniability further complicates matters, since standard PQ signature schemes can produce transferable proofs of communication, conflicting with Signal’s goals.

Additionally, integrating PQ primitives affects message size, authentication binding, and protocol state management, especially when replacing X3DH while maintaining compatibility with the Double Ratchet’s key derivation chains. Any PQ redesign must simultaneously satisfy quantum resistance, reuse safety, authentication guarantees, and asynchronous usability—making the transition a delicate cryptographic and systems-level balancing act rather than a straightforward algorithm upgrade.

\section{Conclusion}

This survey examined nine protocols that collectively secure the majority of Internet communication and found the post-quantum transition at varying stages of maturity. Hybrid key exchange using compact KEMs like ML-KEM has reached production deployment in TLS and Signal, demonstrating that post-quantum confidentiality is achievable today with minimal performance penalty. However, authentication remains a harder problem across the board ---post-quantum signatures are larger, certificate chains inflate substantially, and existing trust infrastructures from the Web PKI to DNSSEC and RPKI were not designed to accommodate them. Protocols that are tightly constrained by message size, such as DNSSEC and BGP, face structural obstacles that may require architectural changes rather than algorithm substitution alone. Meanwhile, application-layer protocols like OpenID Connect introduce a distinct challenge: their security depends not only on transport protection but on the integrity of signed tokens and federated trust models that must themselves be migrated. With NIST targeting deprecation of quantum-vulnerable algorithms by 2030 and full retirement by 2035, the window for planning and executing this transition is narrowing. Organisations should begin now by inventorying their cryptographic dependencies, deploying hybrid modes where available, and tracking the standardisation efforts that will determine how each protocol ultimately achieves quantum resistance.

\bibliographystyle{acm}
\bibliography{bib/main, bib/bgp_rpki_cite, bib/dnssec_cite, bib/ssh_cite, bib/quic_cite, bib/signal_cite, bib/tls_cite, bib/ipsec_cite, bib/pq_class, bib/oidc_cite}

\end{document}